\documentclass[conference]{IEEEtran}

\usepackage{algorithmic}
\usepackage{amsmath}
\usepackage{amssymb}
\usepackage{color}
\usepackage{graphicx}
\usepackage{subcaption}
\usepackage{url}

\sloppypar

\begin{document}

\title{Performance and Optimization Abstractions for Large Scale
  Heterogeneous Systems in the Cactus/Chemora Framework}

\author{
  \IEEEauthorblockN{Erik Schnetter}
  \IEEEauthorblockA{Perimeter Institute for Theoretical Physics,
    Waterloo, Ontario, Canada\\
    Department of Physics, University of Guelph, Guelph, Ontario,
    Canada\\
    Center for Computation \& Technology, Louisiana State University,
    Baton Rouge, Louisiana, USA\\
    Homepage: \url{http://www.perimeterinstitute.ca/personal/eschnetter/}}
}

\date{July 15, 2013}

\maketitle

\begin{abstract}
  We describe a set of lower-level abstractions to improve performance
  on modern large scale heterogeneous systems. These provide portable
  access to system- and hardware-dependent features, automatically
  apply dynamic optimizations at run time, and target
  stencil-based codes used in finite differencing, finite volume,
  or block-structured adaptive mesh refinement codes.
  
  These abstractions include a novel data structure to manage
  refinement information for block-structured adaptive mesh
  refinement, an iterator mechanism to efficiently traverse
  multi-dimensional arrays in stencil-based codes, and a portable API
  and implementation for explicit SIMD vectorization.
  
  These abstractions can either be employed manually, or be targeted
  by automated code generation, or be used via support libraries by
  compilers during code generation. The implementations described
  below are available in the Cactus framework, and are used e.g. in
  the Einstein Toolkit for relativistic astrophysics simulations.
\end{abstract}

\section{Introduction}

Cactus \cite{Goodale:2002a, Cactuscode:web} is a software framework
for high performance computing, notably used e.g. in the Einstein
Toolkit \cite{Loffler:2011ay, EinsteinToolkit:web} for relativistic
astrophysics. The Chemora project \cite{Blazewicz2011}
 aims at significantly
simplifying the steps necessary to move from a physics model to an
efficient implementation on modern hardware. Starting from a set of
partial differential equations expressed in a high level language, it
automatically generates highly optimized code suitable for parallel
execution on heterogeneous systems. The generated code is portable to
many operating systems, and adopts widely used parallel programming
standards and programming models (MPI, OpenMP, SIMD Vectorization,
CUDA, OpenCL).

In this paper, we describe a set of lower-level abstractions available
in the Cactus framework, and onto which Chemora is building. These
abstractions are used by many Cactus components outside the Chemora
project as well.

These abstractions are:
\begin{enumerate}
  \item a novel data structure to manage refinement information for
    block-structured adaptive mesh refinement (section
    \ref{sec:bboxset}),
\item an iterator mechanism to efficiently traverse multi-dimensional
  arrays in stencil-based codes, employing dynamic auto-tuning at run
  time (section \ref{sec:loopcontrol}),
\item a portable API and implementation for explicit SIMD
  vectorization, including operations necessary
  for stencil-based kernels (section \ref{sec:vectorization}).
\end{enumerate}

These abstractions address issues we encountered when porting
Cactus-based applications to modern HPC systems such as Blue Waters
(NCSA), Hopper (NERSC), Kraken (NICS), Mira (ALCF), or Stampede
(TACC). Of course, these abstractions also improve performance on
``regular'' HPC systems, workstations, or laptops.

Below, we describe each of these abstractions in turn, and
conclude with general observations and remarks.

\section{Efficient Bounding Box Algebra}
\label{sec:bboxset}

When using adaptive mesh refinement (AMR), one needs to specify which
regions of a grid need to be refined. The shape of these regions can be
highly irregular. Some AMR algorithms (called cell-based AMR) allow
this decision to be made independently for every cell, others (called
block-structured AMR) require that refined points be clustered into
non-overlapping, rectangular regions for improved efficiency
\cite{Berger1984, bergerrigoutsos}. These
regions can then efficiently be represented e.g. via Fortran-style
arrays on which loop kernels operate. While cell-based AMR
algorithms require tree data structures to represent the refinement
hierarchy, block-structured AMR algorithms (such as available in
Carpet \cite{Schnetter:2003rb, Schnetter:2006pg, CarpetCode:web})
require data structures to represent \emph{sets of bounding
  boxes} describing the regions that make up a particular refinement
level.

A \emph{bounding box} (\verb+bbox+) describes the location and shape
of a rectangular region, a \emph{bounding box set} (\verb+bboxset+)
describes a set of non-overlapping bounding boxes.
There is a direct connection between a bboxset and how data for grid
points are stored in memory.
While a bboxset
can in principle describe any set of grid points (that may have
arbitrary shape and may be disconnected), one assumes that a bboxset
comprises just a few rectangular regions, which will then be handled more
efficiently.

Since a bboxset is used to describe the grid points that make up a
particular refinement level, its points lie on a uniform
grid; see figure \ref{fig:ell} below for an example.
Each grid point can be described by its location, which
can be expressed as $x_0^i + n^i \cdot \Delta x^i$ where $x_0^i$ and
$\Delta x^i$ describe origin and spacing of the grid, and $n^i$
is a vector with integer elements. (The abstract index $i$ denotes
that these are vectors, where $i \in [1 \ldots D]$ in $D$ dimensions.)

Carpet not only uses bounding box sets to describe refined regions, it
also offers a full algebra for bounding box sets. This includes
operations such as set union, intersection, difference, complement,
etc., and also includes additional operations enabled by the grid
structure such as \verb+shift+ (to move a set by a certain offset) or
\verb+expand+ (to grow a set in some directions), or to change the
grid spacing. It is also possible to convert a
bboxset into a normalized list of bboxes.

This full set algebra allows using bboxsets as a convenient base
for implementing many other operations, such as
determining the AMR operators for prolongation, boundary
prolongation, restriction, or refluxing; distributing a refinement
level's grid points onto MPI processes; determining the communication
schedule; or performing consistency checks in a simple-to-express
manner. The price one has to pay is that this requires an efficient
data structure for these operations, such as we describe below.

\subsection{Background}

There exist two simple approaches to
describe sets: one can either enumerate the elements of the set (e.g.
in a list or a tree), or one can view the set as a mapping from
elements to a boolean (storing one boolean for each element e.g. in an
array or a map). The former is efficient if the sets contain few
elements and if the elements can be meaningfully ordered, the latter
is efficient if the number of possible set elements is small.

Unfortunately, neither is the case here: We intend to handle a
refinement level as a set of rectangular bboxes for efficiency;
building a data
structure that disregards this structure and manages points
individually will be much less efficient. At the same time, the
\emph{possible} number of points can be many orders of magnitude
larger than the \emph{actual} number of points in a region. Thus neither
enumerating the grid points making up a bboxset (e.g. via their
integer coordinates) makes sense, nor using a boolean array to
describe which points belong to a refinement.
A more complex data structure is needed.

The literature describes a host of data structures for holding sets of
points, or
to describe sets of regions. For example, GIS (Geographic Information
Systems) heavily rely on such data structures, and $R$-trees or
$R^*$-trees \cite{wiki:rstar-tree} find applications there. While it
would be possible to design an efficient bboxset data structure
based on these, they do not quite fit our problem
description: They assume that the points making up the set are
unstructured (i.e. do not need to be located on a uniform grid), and make no
attempt to cluster these points into bboxes. On the other hand,
$R^*$-trees are able to handle regions with varying point density,
which is not relevant for a uniform grid.

Other block-structured AMR packages introduce data structure to handle
sets of points on uniform grids, but do not provide a full set
algebra. Internally, these bboxsets are often represented as a list of
bounding boxes. For example, AMROC \cite{amrocweb} calls this
structure \verb+BBoxList+. (AMROC is a successor of DAGH, which is in
turn the intellectual predecessor of Carpet.)
Efficient operations may include creating a bboxset from a list of
non-overlapping bboxes, while adding an individual bbox to an existing
bboxset may not be an efficient operation. Specific operations required
for an AMR algorithm are then implemented efficiently, but other
operations -- such as calculating the intersection between two
bboxsets -- are not. Most AMR operations acting on bboxsets are then
implemented in an ad-hoc fashion and may introduce arbitrary
restrictions, e.g. regarding the size of the individual bboxes, or the
number of ghost zones for inter-process communication.

Carpet's previous bboxset data structure was based on a list of
non-overlapping bboxes. It did provide a full algebra of set
operations, but with reduced efficiency. For example, the list of
bboxes was kept normalized, requiring an $O(n^2)$ normalization step after
each set operation, where $n$ is the number of bboxes in the list.
Many other set
operations also had an $O(n^2)$ cost, as is common for set
implementations based on lists. This cost was acceptable for small
numbers of bboxes (say, less than 1,000), but began to dominate the
grid setup time when using more than 1,000 MPI processes, as the
regions owned by MPI processes are described by bboxes.

An earlier attempt to improve the efficiency of bboxsets is
described in \cite{Zebrowski:2011bl}. Unfortunately, this work never
left the demonstration stage.

We are not aware of other literature or source code describing a generic,
efficient data structure to handle sets of points lying on a uniform
grid. To our knowledge, this is a novel data structure for AMR
applications.

\subsection{Discrete Derivatives of Bounding Box Sets}

\begin{figure}
  \centering
  \begin{subfigure}[t]{0.27\linewidth}
    \centering
    \includegraphics[width=\linewidth]{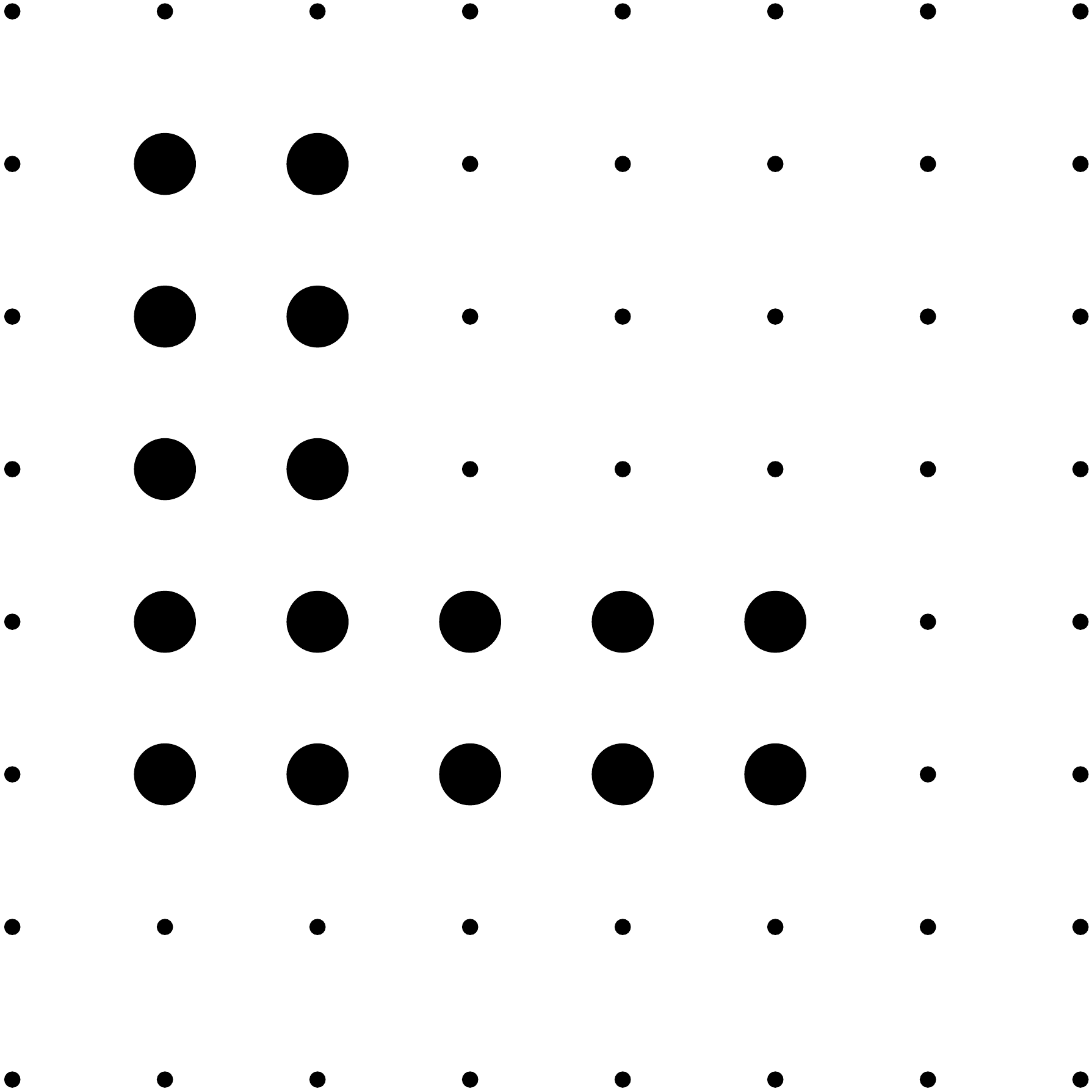}
    \caption{L-shaped region}
    \label{fig:ell}
  \end{subfigure}
  \hfill
  \begin{subfigure}[t]{0.27\linewidth}
    \centering
    \includegraphics[width=\linewidth]{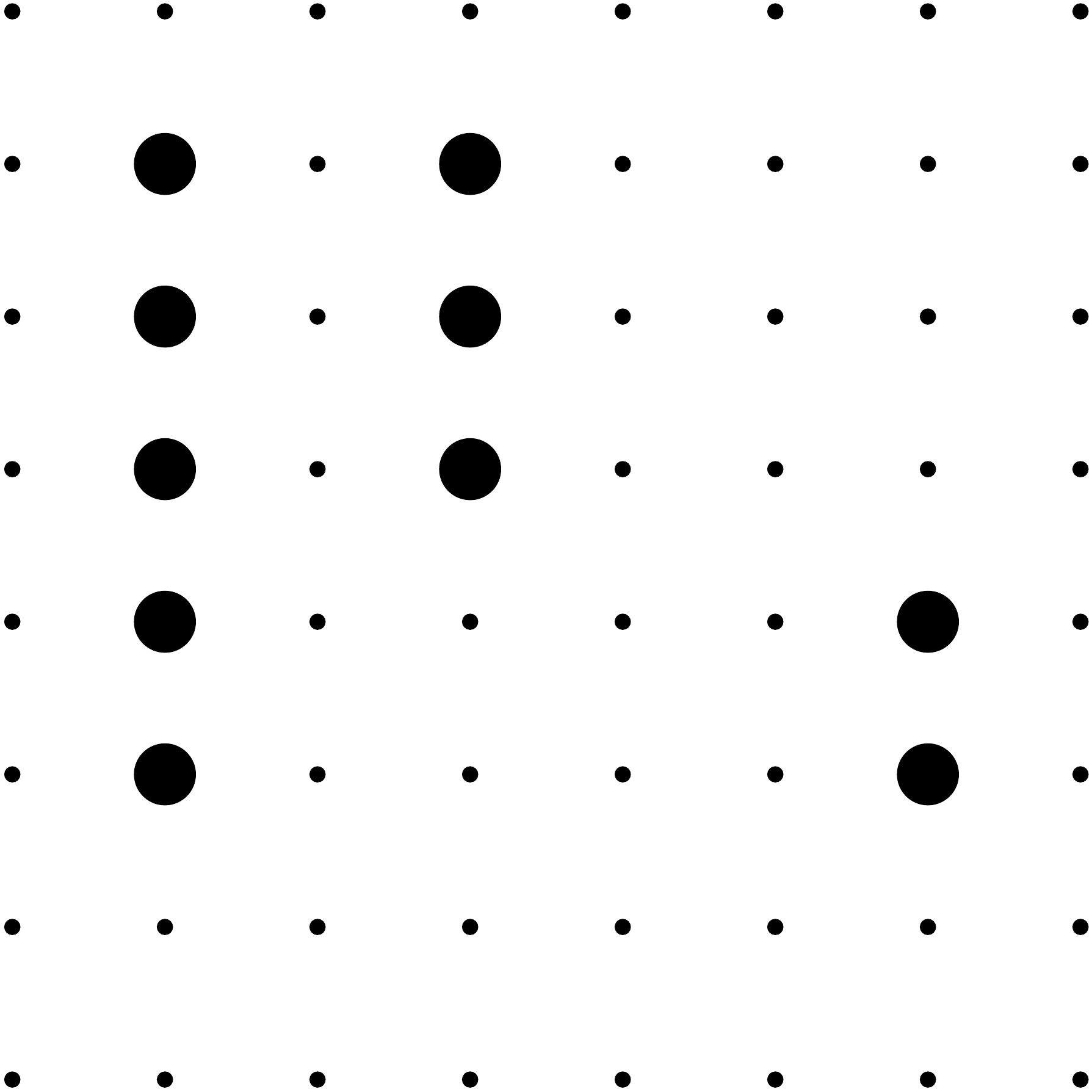}
    \caption{$x$-derivative of this region}
    \label{fig:ell-dx}
  \end{subfigure}
  \hfill
  \begin{subfigure}[t]{0.27\linewidth}
    \centering
    \includegraphics[width=\linewidth]{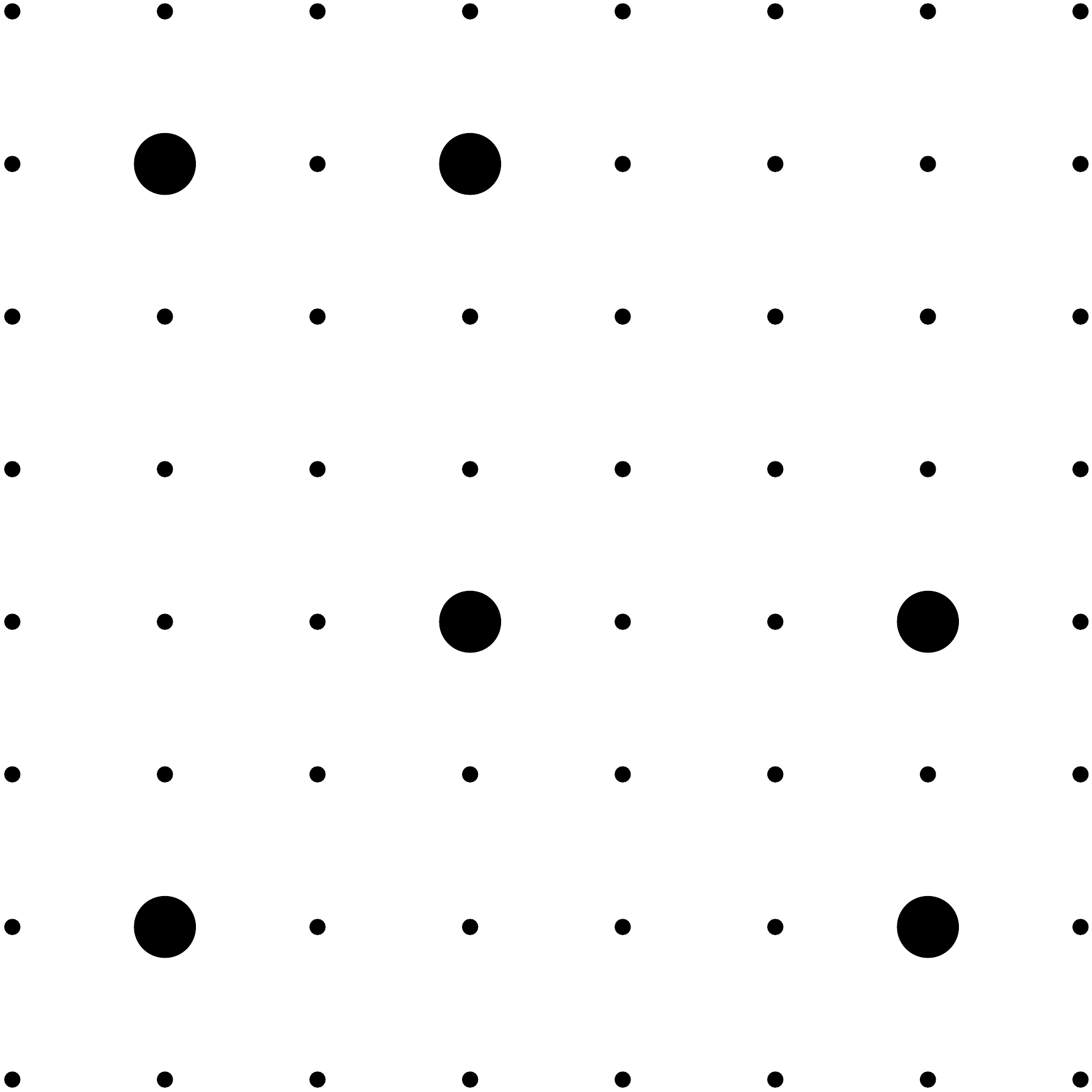}
    \caption{$xy$-derivative of this region}
    \label{fig:ell-dxy}
  \end{subfigure}
  \caption{An L-shaped region and its derivatives. The $xy$-derivative
    consists only of the ``key points'' of this shape, and can be
    stored very efficiently.}
  \label{fig:derivatives}
\end{figure}

Our data structure is based on storing the \emph{discrete derivative}
of a bboxset. An example is shown in figure \ref{fig:derivatives}.
Algebraically, the discrete derivative in the $i$-direction of a
bboxset $R$ is given by
\begin{eqnarray}
  \partial_i R & := & R\; \veebar\; \mathrm{shift}(R, -e^i)
\end{eqnarray}
where $\veebar$ is the symmetric set difference (exclusive or),
$\mathrm{shift}(R, v)$ shifts the bboxset $R$ by a certain offset $v$,
and $e^i$ is the unit vector in direction $i$ ($i$th component is 1,
all other components are 0).

This discrete derivative is equivalent to a finite difference
derivative, applied to the boolean values describing the set
interpreted as integer values modulo 2, i.e. using the
following arithmetic rules for addition: $0 \oplus 0 = 0$, $0 \oplus 1
= 1$, $1 \oplus 1 = 0$.

Note that we choose to use a leftward finite difference; this is a
convention only and has no other relevance. We also note that these
discrete derivatives commute, so that the order in which they are
applied does not matter. Given a set derivative, the anti-derivative
is uniquely defined, and the original set can be readily
recovered:
\begin{eqnarray}
  R & := & \partial_i R\; \veebar\; \mathrm{shift}(R, -e^i)
\end{eqnarray}
This implies that the anti-derivative should be calculated by scanning
from left to right.

The salient point about taking the derivative is that it reduces the
number of elements in a set, assuming that the set has a ``regular''
structure. For example, in two dimensions (as shown
above), an L-shaped region is described by just six points, and in
three dimensions, a cuboid (``3D rectangle'') is described by just eight
points. In fact, the number of points in the derivative of a set
increases with the number of bboxes required to describe it -- this is
exactly the property we are looking for, as the efficiency of a
block-structured AMR algorithm already depends on the number of bboxes
required to represent the bboxset.

Since the number of elements in the derivative of a set is small, we store
these points (i.e. their locations) directly in a tree structure.

Instead of taking derivatives to identify boundaries, one could also
use a run-length encoding; the resulting algorithm would be very
similar.

\subsection{Implementation}

We now describe an efficient algorithm for set operations based on
storing bboxsets as derivatives.

Most set
operations cannot directly be applied to bboxsets stored as
derivatives.
The notable exception for this is the symmetric difference, which can
directly be applied to derivatives:
\begin{eqnarray}
  \label{eq:xor}
  R \veebar S & = & \partial R \veebar \partial S
\end{eqnarray}
where we introduce the notation $\partial R$ to denote subsequent
derivatives in all direction, i.e. $\partial R := \partial_0
\partial_1 R$ in two dimensions, and $\partial R := \partial_0
\partial_1 \partial_2 R$ in three dimensions.
Property (\ref{eq:xor}) follows directly
from the definition of the derivative above and the properties of the
exclusive-or operator.

To efficiently reconstruct a bboxset from its derivative, we employ a
\emph{sweeping algorithm} \cite{wiki:sweeping}.
Instead of
directly taking the derivative of a bboxset $R$ in all directions, we
employ dimensional recursion. We represent a $d$-dimensional bboxset
by taking its derivative in direction $d$, and storing the resulting
set of $d-1$-dimensional bboxsets in a tree structure. We do this
recursively, until we arrive at $0$-dimensional bboxsets. These are
single points, corresponding to a single boolean value that we store
directly, ending the recursion.

Since this data structure represents a bboxset, it is irrelevant how
the bboxset is represented internally. In particular, from the
$d$-dimensional bboxset's representation, it does not matter how the
$d-1$-dimensional bboxsets are internally represented, and from an
algorithm design point of view, the $d-1$-dimensional bboxsets are
directly available for processing.

We now describe how to efficiently evaluate the result of a set
operation acting on two bboxsets $R$ and $S$, calculating $T := R
\odot S$ for an arbitrary set operation $\odot$.
The main idea is to sweep the domain in direction $d$,
keeping track of of the \emph{current state} of the $d-1$-dimensional
subsets $R_{d-1}$, $S_{d-1}$, and $T_{d-1}$ on the sweep line.
(This ``line'' is a $d-1$-dimensional hypersurface in general.)
As the sweep line progresses, we update $R_{d-1}$ and $S_{d-1}$ by
calculating the anti-derivative from our stored derivatives, calculate
$T_{d-1} := R_{d-1} \odot S_{d-1}$, and then calculate and store the
derivate of $T_{d-1}$ in a new bboxset structure.%

The operation $T_{d-1} := R_{d-1} \odot S_{d-1}$ needs to be
re-evaluated whenever $R_{d-1}$ or $S_{d-1}$ change, i.e. once for
each element in the stored derivatives of $R_d$ and $S_d$.
Figure \ref{fig:bboxset-traverse} lists the respective algorithm.


\begin{figure}
  \begin{algorithmic}
    \STMT $R_{d-1} := \{\}$
    \STMT $S_{d-1} := \{\}$
    \STMT $T_{d-1} := \{\}$
    \STMT $n := 0$
    \WHILE{find next $n$ for which $\partial_dR_d$ or $\partial_dS_d$}
    \IF{$\partial_dR_d$ contains an element at $n$}
    \STMT $R_{d_1} := R_{d-1} \veebar \partial_dR_d[n]$
    \ENDIF
    \IF{$\partial_dS_d$ contains an element at $n$}
    \STMT $S_{d_1} := S_{d-1} \veebar \partial_dS_d[n]$
    \ENDIF
    \STMT $T'_{d-1} := T_{d-1}$
    \STMT $T_{d-1} := R_{d-1} \odot S_{d-1}$
    \STMT $\partial_dT_d[n] := T_{d-1} \veebar T'_{d-1}$
    \IF{$\partial_dT_d[n]$ not empty}
    \STMT store $\partial_dT_d[n]$
    \ENDIF
    \ENDWHILE
  \end{algorithmic}
  \caption{Algorithm for traversing two bboxsets $R$ and $S$,
    calculating $T := R \odot S$ where $\odot$ is an arbitrary set
    operation. This algorithm applies to a
    $d$-dimensional set, recursing to $d-1$ dimensions.}
  \label{fig:bboxset-traverse}
\end{figure}

\begin{figure}
  \centering
  \begin{subfigure}{\linewidth}
    \centering
    \vspace{-0.7cm}
    \includegraphics[width=\linewidth]{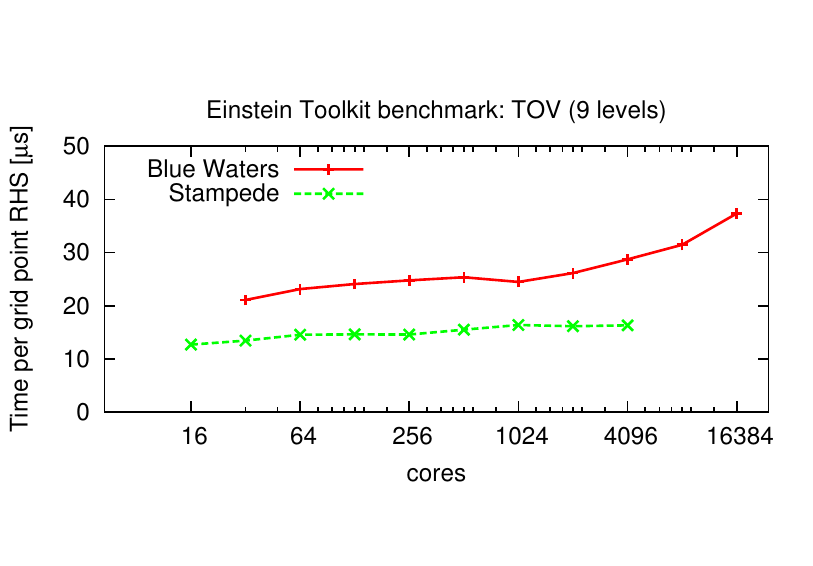}
    \vspace{-1.4cm}
  \end{subfigure}
  \begin{subfigure}{\linewidth}
    \centering
    \vspace{-0.7cm}
    \includegraphics[width=\linewidth]{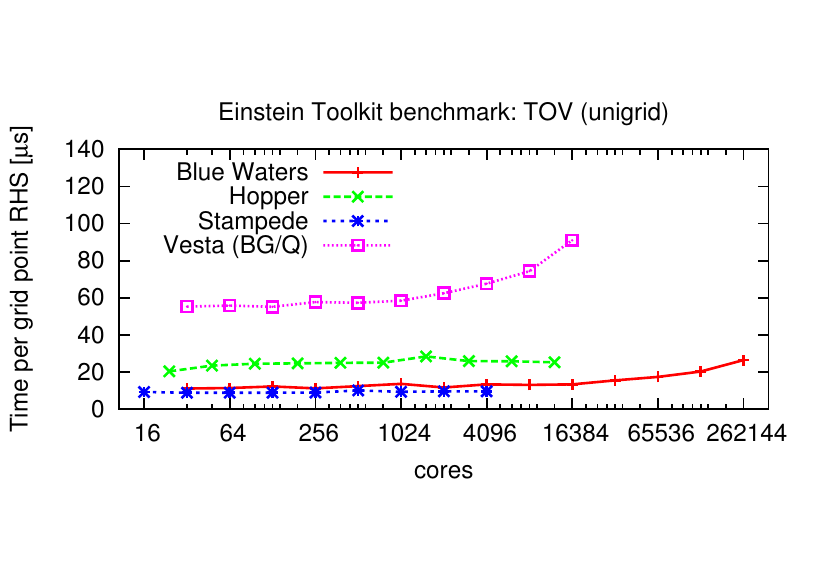}
    \vspace{-1.3cm}
  \end{subfigure}
  \caption{Weak scaling benchmarks for a relativistic astrophysics
    application with Carpet, using nine refinement levels (top) and a
    uniform grid (bottom). Smaller times are better, ideal scaling is
    a horizontal line. Carpet's AMR implementation scales to 10k+
    cores. With a uniform grid, Carpet scales to 250k+ cores.}
  \label{fig:bench}
\end{figure}

Given that accessing set elements stored in a tree has a cost of
$O(\log n)$, set operations implemented via the algorithm above have a
cost that can be bounded by $O([n_d \log n_d]^d)$, where $d$ is the
number of dimensions, and $n_d$ is the maximum number of bboxes
encountered by a scan line in direction $d$. In non-pathological
cases, $n_d \approx n^{1/d}$, leading to a log-linear cost. Figure
\ref{fig:bench} demonstrates then scalability of Carpet for a weak
scaling benchmark, when this bboxset algorithm is used for all set
operations.

\subsection{Future Work}

This derivative bboxset data structure and its
associated algorithms are serial, as the sweeping algorithm is sequential
and does not lead to a natural parallelization. (Of course, different
sets can still be processed in parallel.)

One parallelization approach would be
to break each bboxset into several independent pieces and to process
these in parallel. This would then also require stitching the
results together after each set operation.

\section{Dynamic Loop Optimizations}
\label{sec:loopcontrol}

Most CPUs and accelerators (if present) of modern HPC
systems are multi-core systems with a deep memory hierarchy, where
each core requires SIMD vectorization to obtain the highest
performance. In addition, in-order-execution systems (i.e.
accelerators, including Blue Gene/Q) require SMT (symmetric
multi-threading) to hide memory access and instruction latencies.
These architectures require significant programmer effort to achieve
good single-node performance, even when leaving distributed memory MPI
programming aside.%
\footnote{Single-\emph{core} performance is not really relevant here,
  since (a) applications will use more than one core per node, and (b)
  the individual cores interact at run time e.g. via cache access
  patterns.}

Ignoring the issues of SIMD vectorization here (see section
\ref{sec:vectorization} below), one would hope that language standards
and implementations such as OpenMP or OpenCL allow programmers
to ensure efficiency. However, this is not so, for several reasons:
\begin{itemize}
  \item neither OpenMP nor OpenCL allow distinguishing between SMT,
    where threads share all caches, and coarse-grained multi-threading,
    where most cache levels are not shared;
  \item it is very difficult, if not impossible to reliably predict
    performance of compute kernels, so that dynamic (run-time)
    decisions regarding optimizations are necessary;
  \item the most efficient multi-threading algorithms need to be aware
    of cache
    line boundaries, which also needs to factor into how
    multi-dimensional arrays are allocated; neither OpenMP nor OpenCL
    provide support for this.
\end{itemize}

Here we present \emph{LoopControl}, an iterator mechanism to
efficiently loop over multi-dimensional arrays for stencil-based loop
kernels. LoopControl parallelizes iterations across multiple
threads, across multiple SMT threads, performs loop
tiling to improve cache efficiency, and honours SIMD vector sizes to
ensure an efficient SIMD vectorization is possible.

LoopControl monitors the performance of
its loops, and dynamically adjusts its parameters to improve
performance. This not only immediately adapts to different machines
and to code
modifications, but also to differing conditions at run time such as
changes to array sizes (e.g. due to AMR) or changes to the physics
behaviour in loop kernels. LoopControl uses a \emph{random-restart
  hill-climbing} algorithm for this dynamic optimization.

The multi-threading is based on OpenMP threads, but employs a dynamic
region selection and load distribution mechanism to handle kernels
with non-uniform cost per iteration.

LoopControl employs \emph{hwloc} \cite{hwlocweb} to query the system
hardware, and also queries MPI and OpenMP about process/thread setups.
hwloc also reports thread-to-core and thread-to-cache
bindings that are relevant for performance. All information is
gathered automatically, requiring no user setup to achieve good
performance.

LoopControl dynamically auto-tunes stencil codes at run time. This is
fundamentally different from traditional auto-tuning, which surveys
the parameter space for a set of optimizations ahead of time, and then
re-uses these survey results at run time. See e.g. \cite{Datta2008}
for a description of ahead-of-time auto-tuning of stencil-based codes, or
\cite{Balaprakasha2011} for a description of ahead-of-time
auto-tuning search algorithms.

Ahead-of-time surveys have the disadvantage that they need to be
repeated for each machine on which the code runs, for each compiler
version/optimization setting, for each modification to the loop
kernel, and also for different array sizes. This makes it prohibitively
expensive to use in a code that undergoes rapid development, or where
adaptive features such as AMR are used. LoopControl does not have
these limitations, and to our knowledge, LoopControl's dynamic
auto-tuning algorithm is novel.

\subsection{Loop Traversal}

LoopControl assumes that each loop iteration is independent of the
others, and can thus be executed in parallel or in an arbitrary order.

Most architectures have several levels of caches. LoopControl
implicitly chooses one cache level for which it optimizes. The
random-restart algorithm (see below) will explore optimizing for other cache
levels as well, and will settle for that level that yields the largest
performance gain. It would be straightforward to implement support for
multiple cache levels, but it is not clear that this would
significantly improve
performance in practice.

LoopControl uses the following mechanisms, in this order, to split the
index space of a loop:
\begin{enumerate}
\item coarse-grained (non-SMT) multithreading (expecting no shared
  caches)
\item iterating over loop tiles (each expected to be small enough to
  fit into the cache)
\item iterating within loop tiles
\item fine-grained (SMT) multithreading (expecting to share the finest
  cache level)
\item SIMD vectorization (see section \ref{sec:vectorization} below).
\end{enumerate}

The index space is only known at run time. It is split multiple times
to find respective smaller index spaces for each of the mechanisms
described above. Each index space is an integer multiple of the next
smaller index space, up to boundary effects.

Certain index space sizes and offsets have to obey certain constraints:
\begin{itemize}
  \item SIMD vectorization requires that its index space to be aligned with
    and have the same size as the SIMD hardware vector size.
  \item The SMT multithreading index space should be a multiple of the
    vector size, so that partial vector store operations are not
    required except at loop boundaries (as some hardware does not offer
    thread-safe partial vector stores).
  \item The number of SMT and non-SMT threads is determined by the
    operating system upon program start, and are not modified (i.e.
    all threads are used).
  \item Loop tiles should be aligned with cache lines for efficiency.
\end{itemize}

Since LoopControl cannot influence how arrays are allocated, the
programmer needs to specify the array alignment, if any. The first
array element is expected be aligned with the vector size or the
cache line size (which can always be ensured when the array is
allocated), and the array dimensions may or may not be padded to
multiples of the vector size or cache line size. Higher alignment leads
to more efficient execution on some hardware, since edge effects such
as partial vector stores or partial cache line writes can be avoided.
LoopControl offers support for all cases.

\subsection{Random-Restart Hill-Climbing}

Since each loop behaves differently (obviously), and since this
performance behaviour also depends on the loop bounds, LoopControl
optimizes each \emph{loop setup} independently. A loop setup includes
the loop's source code location, index space, array alignment, and number of
threads available.

Several execution parameters describe how a loop setup is executed,
describing how the index space is split according to
the mechanisms described above.

Each newly encountered loop setup has its initial execution parameters
chosen heuristically. As timing measurements of the loop setup's
execution become available, these parameter settings are optimized. It
is well known that the execution time of a loop kernel depends on
optimization parameters in a highly non-linear and irregular manner,
with many threshold effects present.
Simplistic optimization algorithms will thus fail. For this
optimization, we use a random-restart hill-climbing algorithm
as described in the following.

Our optimization algorithm has two competing goals: (1) for a given
execution parameter setting, quickly find the nearby local optimum,
and (2) do not get stuck in local optima; instead, explore the whole
parameter space. To find a local optimum, we use a \emph{hill
  climbing} algorithm: we explore the local neighbourhood of a given
parameter setting, and move to any setting that leads to a shorter run
time, discarding parameter settings that lead to longer execution
times. To explore the whole parameter space, we use a \emph{random
  restart} method: once we arrived in a local optimum, we decide with
a certain, small probability to chose a random new parameter setting. After
exploring the neighbourhood of this new parameter setting, we either
remain there (if it is better), or return to the currently known
best setting.

There is one major difference between an ahead-of-time exploration of
the parameter space, and a dynamic optimization at run time: The goal
of an ahead-of-time exploration is to find the best possible parameter
setting, while the goal of a run-time optimization is to reduce the
overall run time. A bad parameter setting can be significantly worse
than a mediocre parameter setting, and can easily have a running time that
is an order of magnitude higher. That means that exploring even one
such bad parameter setting has a cost that is only amortized if one
executes hundreds of loops with good parameter settings.

This makes it important to be cautious about exploring the
parameter space, and to very quickly abort any excursion that
significantly worsens the run time. It is much more important to find
a mediocre improvement and to find it quickly, than to find the optimum
parameter choice and incurring a large overhead. In
particular, we find that genetic algorithms or simulated annealing
spend much too much time on bad parameter settings, and while they may
ultimately find good parameter settings, this comes at too great a
cost to be useful for a dynamic optimization to be applied at
run-time.

For the relatively large kernels present in our
astrophysics application, we observe roughly a 10\% improvement over a
naive OpenMP parallelization via \verb+#pragma omp parallel for+
for the first loop executions via our heuristic parameter choices, and
an additional approximately 10\% improvement in the long run via
LoopControl's dynamic optimizations.

\subsection{Future Work}

It may be worthwhile to save and restore execution parameter settings
and their respective timings. Although these may be invalidated by
code modifications or changes to the build setup, this would provide a
way to remember exceptionally good parameter settings that may
otherwise be difficult to re-discover.

In particular in conjunction with OpenCL, where it is simple to
dynamically re-compile a loop kernel, LoopControl's optimizations could
also include compile-time parameter settings such as loop unrolling or
prefetching.

In addition to these low-level loop execution optimizations, one can
also introduce optimizations at a higher level, such as e.g. loop
fission or loop fusion. These optimizations can have large impacts on
performance if they make code fit into instruction- or data-caches.
Combining LoopControl's optimizer with a way to select between
different (sets of) loop kernels would be straightforward.

\section{SIMD Vectorization}
\label{sec:vectorization}

Modern CPUs offer SIMD (short vector) instructions that operate on a
small number of floating point values simultaneously; the exact number
(e.g. 2, 4, or 8) is determined by the hardware
architecture.
To achieve good performance, it is essential that SIMD
instructions are used when compiling compute kernels; not doing so
will generally reduce the possible theoretical peak performance by
this factor. Of course, this is relevant only for compute-bound
kernels.

\subsection{Background}

However, using SIMD instructions typically comes with a set of
restrictions that need to be satisfied; if not, SIMD instructions
either cannot be used, or lose a significant fraction of their
performance. One of these restrictions is that it is efficient to
perform element-wise operations, but quite inefficient to reduce
across a vector. That is, while e.g. $a_i:=b_i+c_i$ is highly
efficient, the operation $s:=\sum_i a_i$ will be relatively expensive.
This means that one should aim to vectorize across calculations that
are mutually independent. As a rule of thumb, it is better to
vectorize across different loop iterations than to try and find
independent operations within a single iteration.

Another restriction concerns memory access patterns. Memory and cache
subsystems are these days highly vectorized themselves (with
typical vector sizes of e.g. 64 bytes), and efficient
load/store operations for SIMD vectors require that these vectors are
\emph{aligned} in memory. Usually, a SIMD vector with a size of $N$
bytes needs to be located at an address that is a multiple of $N$.
Unaligned memory accesses are either slower, or are not possible at
all and then need to be split into two aligned memory accesses and
shift operations.

Finally, if one vectorizes across loop iterations, the number of
iterations may not be a multiple of the vector size. Similarly, if one
accesses an array in a loop, then the first accessed array element may
not be aligned with the vector size. In both cases, one needs to
perform operations involving only a subset of the elements of an SIMD
vector. This is known as \emph{masking} the vector operations. The
alternative -- using scalar operations for these edge cases --
is very expensive if the vector size is large.

Unfortunately, the programming languages that are widely used in HPC today
(C, C++, Fortran) do not offer any constructs that would directly map
to these SIMD machine instructions, nor do they offer declarations
that would ensure the necessary alignment of data structure. It is
left to the compiler to identify kernels where SIMD instructions can
be used to increase efficiency, and to determine whether data
structures have the necessary alignment.
Often,
system-dependent source code annotations can be used to help the
compiler, such as e.g. \verb+#pragma ivdep+ or \verb+#pragma simd+ for
loops, or \verb+__builtin_assume_aligned+ for pointers.

Generally, compiler-based vectorization works fine for small loop
kernels, surrounded by simple loop constructs, contained in small
functions. This simplifies the task of analyzing the code, proving
that vectorization does not alter the meaning, and allowing estimating
the performance of the generated code to ensure that vectorization
provides a benefit. However, we find that the converse is also true:
large compute kernels, kernels containing non-trivial control flow
(if statements), or using non-trivial math functions (exp, log) will
simply not be vectorized by a given compiler. ``Convincing'' a certain
compiler that a loop should be vectorized remains a highly
system-specific and vendor-specific (i.e. non-portable) task. In
addition, if a loop is vectorized, then the generated code may make
pessimistic assumptions regarding memory alignment that lead to
sub-ideal performance, in particular when stencil operations in
multi-dimensional arrays are involved.

The root of the problem seems to be that the compiler's optimizer
does not have access to sufficiently rich, high-level
information about the employed algorithms and their implementation to
make good decisions regarding vectorization. (The same often holds true for
other optimizations as well, such as e.g. loop fission/fusion, or
cloning functions to modify their interfaces.) We hope that the coming
years will lead to widely accepted ways to pass such information to
the compiler, either via new languages or via source code annotations.
For example, the upcoming OpenMP 4.0 standard will provide a
\verb+#pragma omp simd+ to enforce vectorization, GCC is already
providing \verb+__builtin_assume_aligned+ for pointers, and Clang's
vectorizer has as of version 3.3 arguably surpassed that of GCC 4.8,
justifying our hope that things are improving.

\subsection{Manual Vectorization}

The hope for future compiler features expressed in the previous
section does not help performance today. Today, vectorizing a
non-trivial code requires using architecture-specific and sometimes
compiler-specific \emph{intrinsics} that provide C/C++ datatypes and
function calls mapping directly to respective vector types and vector
instructions that
are directly supported by the hardware.
This allows
achieving very high performance, at the cost of portability.

For example, the simple loop

\begin{small}
\begin{verbatim}
for (int i=0; i<N; ++i) {
  a[i] = b[i] * c[i] + d[i];
}
\end{verbatim}
\end{small}

\noindent can be manually vectorized with Intel/AMD's SSE2 intrinsics
(for all 64-bit Intel and AMD CPUs) as

\begin{small}
\begin{verbatim}
#include <emmintrin.h>
for (int i=0; i<N; i+=2) {
  __m128d ai, bi, ci, di;
  bi = _mm_load_pd(&b[i]);
  ci = _mm_load_pd(&c[i]);
  di = _mm_load_pd(&d[i]);
  ai = _mm_add_pd(_mm_mul_pd(bi, ci), ci);
  _mm_store_pd(&a[i], ai);
}
\end{verbatim}
\end{small}

\noindent or with IBM's QPX intrinsics (for the Blue Gene/Q) as

\begin{small}
\begin{verbatim}
#include <builtins.h>
for (int i=0; i<N; i+=4) {
  vector4double ai, bi, ci, di;
  bi = vec_lda(0, &b[i]);
  ci = vec_lda(0, &c[i]);
  di = vec_lda(0, &d[i]);
  ai = vec_madd(bi, ci, di);
  vec_sta(ai, 0, &a[i]);
}
\end{verbatim}
\end{small}

\noindent These vectorized loops assume that the array size $N$ is a
multiple of the vector size, and that the arrays are aligned with the
vector size. If this is not the case, the respective vectorized code
is more complex.

While the \emph{syntax} of the vectorized kernels looks
quite different, the \emph{semantic} transformations applied to the
original kernel are quite similar. Vector values are stored in
variables that have a specific type (\verb+__mm128d+,
\verb+vector4double+), memory access operations have to be denoted
explicitly (\verb+_mm_load_pd+, \verb+vec_lda+), and arithmetic
operations become function calls (\verb+_mm_add_pd+, \verb+vec_madd+).
Other architectures require code transformations along the very same
lines.

Note that QPX intrinsics support a fused multiply-add (\emph{fma})
instruction that calculates $a\cdot b+c$ in a single instruction (and
presumably also in a single cycle). Regular C or C++ code would
express these via separate multiply and add operations, and it would
be the task of the compiler to synthesize such fma operations when
beneficial. When writing vectorized code manually, the compiler will
generally not synthesize vector fma instructions, and this
transformation has to be applied explicitly. Today, most CPU
architectures support fma instructions.

Vector architectures relevant for high-performance computing these
days include Intel's and AMD's SSE instructions, Intel and AMD's AVX
instructions (both SSE and AVX exist in several variants), Intel's Xeon
Phi vector instructions, IBM's Altivec and VSX instructions for Power
CPUs, and IBM's QPX instructions for the Blue Gene/Q. On low-power
devices, ARM's NEON instructions are also important.

\subsection{An API for Explicit Vectorization}

Based on architecture- and compiler-dependent intrinsics, we have
designed and implemented a portable, efficient API for explicit loop
vectorization. This API targets stencil-based loop kernels, as can
e.g. be found in codes using finite differences or finite volumes,
possibly via block-structured adaptive mesh refinement. Our
implementation \verb+LSUThorns/Vectors+ uses C++ and supports all
major current HPC architectures \cite{Loffler:2011ay,
  EinsteinToolkit:web}.

The API is intended to be applied to existing scalar codes in a
relatively straightforward manner. Data structures do not need to be
reorganized, although it may provide a performance benefit if they
are, e.g. ensuring alignment of data accessed by vector instructions,
or choosing integer sizes compatible with the available vector
instructions.

The API consists of the following parts:
\begin{itemize}
\item data types holding vectors of real numbers (float/double),
  integers, and booleans (e.g. for results of comparison operators, or
  for masks);
\item the usual arithmetic operations (+ - * /, copysign, fma, isnan,
  signbit, sqrt, etc.), including comparisons, boolean operations, and
  an if-then construct;
\item ``expensive'' math functions, such as cos, exp, log, sin, etc.
  that are typically not available as hardware instructions;
\item memory load/store operations, supporting both aligned and
  unaligned access, supporting masks, and possibly offering to bypass
  the cache to improve efficiency;
\item helper functions to iterate over a index ranges, generating
  masks, and ensuring efficient array access, suitable in particular
  for stencil-based kernels.
\end{itemize}
We describe these parts in more detail below.

The OpenCL C language already provides all but the last item. Once
mature OpenCL implementations become available for HPC platforms --
that is, for the CPUs on which the applications will be running, not
only for accelerators that may be available there -- this API could be
replaced by programming in OpenCL C instead.
We are actively involved in the \emph{pocl}
(\emph{Portable Computing Language}) project \cite{poclweb} which
develops a portable OpenCL implementation based on the LLVM infrastructure
\cite{llvmweb}.

\subsubsection{Data Types and Arithmetic Operations}

The first two items -- data types and arithmetic operations -- can be
directly mapped to the vector intrinsics available on the particular
architecture. We remark that vectorized integer operations are often
not available, and that vectorized boolean values are internally often
represented and handled quite differently from C or C++.

For each architecture, the available vector instruction set and vector
sizes are determined automatically at compile time, and the most
efficient vector size available is chosen. Both double and single
precision vectors are supported.

For several architectures, this API is implemented via macros instead
of via inline functions. Surprisingly, several widely used compilers
for HPC systems cannot handle inline functions efficiently. The most
prominent consequence of this is that operator overloading is not
possible; instead, arithmetic operations have to be expressed in a
function call syntax such as \verb+vec_add(a,b)+. While
straightforward, this unfortunately reduces readability somewhat.

A trivial implementation, useful e.g. for debugging, maps this API to
scalar operations without any loss of efficiency.

Using our API, the example from above becomes

\begin{small}
\begin{verbatim}
#include <vectors.h>
for (int i=0; i<N; i+=CCTK_REAL_VEC_SIZE) {
  CCTK_REAL_VEC ai, bi, ci, ci;
  bi = vec_load(&b[i]);
  ci = vec_load(&c[i]);
  di = vec_load(&d[i]);
  ai = vec_madd(bi, ci, di);
  vec_store(&a[i], ai);
}
\end{verbatim}
\end{small}

\noindent This code is portable across many architectures. However,
this example still assumes that all arrays are aligned with the vector
size, that the array size is a multiple of the vector size, and does
not include any cache optimizations.

\verb+if+ statements require further attention when vectorizing, since
different elements of a vector may lead to different paths through the
code. Similarly, the logical operators \verb+&&+ and \verb+||+ cannot
have shortcut semantics with vector operands (see e.g. the OpenCL
standard \cite{openclweb}). To translate \verb+if+ statements, we provide a
function \verb+ifthen(cond, then, else)+ with a definition very
similar to the \verb+?:+ operator, but without shortcut semantics.
(This corresponds to the OpenCL \verb+select+ function, except for the
order of the arguments.)

To vectorize an \verb+if+ statement, it needs to be rewritten using this
\verb+ifthen+ function, taking into account that both the \emph{then}
and the \emph{else} branches will be evaluated for all vector
elements. Often, declaring separate local variables for the
\emph{then} and the \emph{else} branches and moving all memory store
operations (if any) out of the \verb+if+ statement (and turning them into
masked store operations if necessary) make this transformation
straightforward.

\subsubsection{``Expensive'' Math Functions}

Some compilers (IBM, Intel) offer efficient implementations of the
``expensive'' math functions that can be used
(\verb+mass_simd+, \verb+mkl_vml+), while other compilers (GCC, Clang)
do not. To support system architectures other than IBM's and Intel's,
we have implemented an open-source library \emph{Vecmathlib}
\cite{vecmathlibweb, Jaaskelainen2013} providing portable, efficient,
vectorized math functions.\footnote{Under some circumstances, this
  library is for scalar code faster than glibc on Intel/AMD CPUs.}

The OpenCL C language standard requires that these math functions be
available for vector types. For the pocl project's OpenCL compiler, we
thus use Vecmathlib to implement these where no vendor library is
available.

\subsubsection{Memory Access}

The API supports a variety of access modes for memory load/store
operations that are likely to occur in
stencil-based codes. In particular, great care has been taken to
ensure that the most efficient code is generated depending on either
compile-time or run-time guarantees that the code can make regarding
alignment. Some code transformations, such as array padding of
multi-dimensional arrays, enable such guarantees and can thus improve
performance.

Let us consider a slightly more complex example using stencil
operations. The code below calculates a derivative via a forward
finite difference:

\begin{small}
\begin{verbatim}
for (int i=0; i<N-1; ++i) {
  a[i] = b[i+1] - b[i];
}
\end{verbatim}
\end{small}

\noindent We assume that the arrays \verb+a+ and \verb+b+ are aligned
with the vector size, and that $N-1$ is a multiple of the vector size.
This code can then be vectorized to

\begin{small}
\begin{verbatim}
#include <vectors.h>
for (int i=0; i<N-1; i+=CCTK_REAL_VEC_SIZE) {
  CCTK_REAL_VEC ai, bi, bip;
  bi  = vec_load(&b[i]);
  bip = vec_loadu_off(+1, &b[i+1]);
  ai  = vec_sub(bip, bi);
  vec_store(&a[i], ai);
}
\end{verbatim}
\end{small}

\noindent Here, the function \verb+vec_loadu_off(offset, ptr)+ loads a
value from memory that is located at an offset of $+1$ from an aligned
value. This specification expects that the offset is known at compile
time, and allows the compiler to generate the most efficient code for
this case. A similar function \verb+vec_loadu(ptr)+ allows loading
unaligned values if the offset is unknown at compile time. Equivalent
functions exist for storing values.

\subsubsection{Iterators}

Finally, our API provides an
``iterator'' to simplify looping over index ranges. Typically, only
the innermost loop of a loop nest is vectorized, and it is expected
that this loop has unit stride. This iterator also sets a mask to
handle edge cases at the beginning and end of the index range.
This is also connected to shared memory parallelization such as via
OpenMP, where one wants to ensure that an OpenMP parallelization of
the innermost loop does not introduce unaligned loop bounds.

The scalar code below evaluates a centered finite difference:

\begin{small}
\begin{verbatim}
for (int i=1; i<N-1; ++i) {
  a[i] = 0.5 * (b[i+1] - b[i-1]);
}
\end{verbatim}
\end{small}

\noindent We assume again that the arrays \verb+a+ and \verb+b+ are
aligned with the vector size. We also assume that the array is padded,
so that we can access elements that are ``slightly'' out of bounds
without causing a segmentation fault. Both conditions can easily be
guaranteed by allocating the arrays correspondingly, e.g. via
\verb+posix_memalign+. If the arrays' alignment is not known at
compile time, then they need to be accessed via \verb+vec_loadu+ and
\verb+vec_storeu+ functions instead. We make no other assumptions, and
the array can have an arbitrary size. This leads to the following
vectorized code:

\begin{small}
\begin{verbatim}
#include <vectors.h>
VEC_ITERATE(i, 1, N-1) {
  CCTK_REAL_VEC ai, bim, bip;
  bim = vec_loadu_off(-1, &b[i-1]);
  bip = vec_loadu_off(+1, &b[i+1]);
  ai  = vec_mul(vec_set1(0.5), vec_sub(bip, bim));
  vec_store_nta_partial(&a[i], ai);
}
\end{verbatim}
\end{small}

\noindent The macro \verb+VEC_ITERATE(i, imin, imax)+
expands to a loop that iterates
the variable \verb+i+ from \verb+imin+ to \verb+imax+ with a stride of
the vector size. It also ensures that \verb+i+ is always a multiple of
the vector size, starting from a lower value than \verb+imin+ if
necessary. Additionally, it prepares an
(implicitly declared) mask in each iteration.

The suffix \verb+_partial+ in the vector
store operation indicates that this mask is taken into account when
storing. The
code is optimized for the case where all vector elements are stored.
The suffix \verb+_nta+ invokes a possible cache optimization, if
available. It indicates that the stored value will in the near future
not be accessed again (``non-temporal access'').
This hint can be used by the implementation to
bypass the cache when storing the value.

Most CPU architectures do not support masking arbitrary vector
operations, while masking load/store operations may be supported. In
the examples given here, we only mask store operations, assuming that
arrays are sufficiently padded for load operations to always succeed. The
unused vector elements are still participating in calculations, but
this does not introduce an overhead.

This iterator provides provides a generic mechanism to traverse arrays
holding scalar values via vectorized operations. It thus provides the
basic framework to enable vectorization for a loop,
corresponding to a \verb+#pragma simd+ statement. By implicitly
providing masks that can be used when storing values, aligned and
padded arrays are handled efficiently.

Different from the previous items, this iterator is applicable even
for OpenCL C code, since no equivalent constructs exist in the language.


\subsection{Applications}

The API described above allows explicitly vectorizing C++ code. While
somewhat tedious, it is in our experience straightforward to vectorize
a large class of scalar codes where vectorization is beneficial.
There is special support for efficient support of stencil-based
codes on block-structured grids using multi-dimensional arrays.

While manual vectorization is possible, this API also lends itself for
automated code generation. We use Kranc \cite{Husa:2004ip, Kranc:web}
to create Cactus components from partial differential equations
written in Mathematica notation, and have modified Kranc's back-end to
emit vectorized code. Mathematica's pattern matching capabilities are
ideal to apply optimizations to the generated vector expressions
that the compiler is unwilling to perform.

\section{Conclusion}

This paper describes a set of abstractions to improve performance on modern
large scale heterogeneous systems targetting stencil-based codes.
These abstractions are available in the Cactus framework, and are used
in ``real-world'' applications, such as in relativistic astrophysics
simulations via the Einstein Toolkit.

Our implementations of these abstractions require access to low-level
system information. Especially hwloc \cite{hwlocweb} and PAPI
\cite{papiweb} provide valuable information. While hwloc is very
portable and easy to use, we are less satisfied with the state of PAPI
installations; these are
often not available (and neither are alternatives),
not even on freshly installed cutting-edge
systems. We are highly dissatisfied with this situation, which
forces us to resort to crude overall timing measurements to evaluate
performance.

While our performance and optimization abstractions are portable, they
are by their very nature
somewhat low-level, and using them directly e.g. from C++ code can be
tedious, although straightforward. We anticipate that they
will see most use either via automated code generation (e.g. via Kranc
\cite{Husa:2004ip, Kranc:web}), or via including them into compiler
support libraries (e.g. via pocl \cite{poclweb}).



%

\section*{Acknowledgements}

We thank Marek Blazewicz, Steve Brandt, Peter Diener, Ian Hinder,
David Koppelman, and Frank L\"offler for valuable discussions and
feedback. We also thank the Cactus and the Einstein Toolkit developer
community for volunteering to test these implementations in their
applications.

This work was supported by NSF award 0725070 \emph{Blue Waters}, NSF
awards 0905046 and 0941653 \emph{PetaCactus}, NSF award 1212401
\emph{Einstein Toolkit}, and an NSERC grant to E. Schnetter.

This work used computational systems at ALCF, NCSA, NERSC, NICS,
Sharcnet, TACC, as well as private systems at Caltech, LSU, and the
Perimeter Institute.

\bibliographystyle{amsplain-url}
\bibliography{abstract,einsteintoolkit,xscale2013}

\end{document}